\documentclass[aps, prb, reprint, superscriptaddress, floatfix]{revtex4-2}
\usepackage{graphicx}
\usepackage{dcolumn}
\usepackage{bm}
\usepackage{physics}
\usepackage{amsfonts}
\usepackage{mathrsfs}
\usepackage{subfigure}
\usepackage[svgnames]{xcolor}
\usepackage[colorlinks=true,linkcolor=NavyBlue,anchorcolor=red,citecolor=NavyBlue, urlcolor=NavyBlue]{hyperref}
\usepackage{color}
\usepackage{xcolor}
\usepackage{ulem}

\DeclareMathAlphabet{\pazocal}{OMS}{zplm}{m}{n}
\usepackage{amsmath}
\usepackage{amssymb}
\usepackage[T1]{fontenc}
\begin{document}

\preprint{APS/123-QED}

\title{Pseudospin Formulation of Quench Dynamics in the Semiclassical Holstein Model}

\author {Lingyu Yang}
\affiliation{Department of Physics, University of Virginia, Charlottesville, Virginia, 22904, USA}

\author {Ho Jang}
\affiliation{Department of Physics, University of Virginia, Charlottesville, Virginia, 22904, USA}

\author{Sankha Subhra Bakshi}
\affiliation{Department of Physics, University of Virginia, Charlottesville, Virginia, 22904, USA}

\author {Yang Yang}
\affiliation{Department of Physics, University of Virginia, Charlottesville, Virginia, 22904, USA}

\author {Gia-Wei Chern}
\affiliation{Department of Physics, University of Virginia, Charlottesville, Virginia, 22904, USA}

\begin{abstract}
We present a pseudospin formulation for the post-quench dynamics of charge-density-wave (CDW) order in the half-filled spinless Holstein model on a square lattice, assuming spatially homogeneous evolution. This Anderson pseudospin description captures the coherent nonequilibrium dynamics of the coupled electron–lattice system. Numerical simulations reveal three distinct dynamical regimes of the CDW order parameter following a quench—locked oscillations, Landau-damped dynamics, and overdamped relaxation—closely paralleling quench dynamics in BCS superconductors and other electronically driven symmetry-breaking phases. Crucially, however, the presence of dynamical lattice degrees of freedom leads to qualitatively different long-time behavior. In particular, while the oscillation amplitude is reduced in the damped regimes, CDW oscillations do not fully decay but instead persist indefinitely due to feedback from the lattice field. We further show that these persistent oscillations are characterized by a nonequilibrium electronic distribution, which provides an intuitive understanding of both their amplitude and the renormalization of the oscillation frequency relative to the bare Holstein phonon frequency. Our results highlight the essential role of lattice dynamics in nonequilibrium ordered phases and establish a clear distinction between electron–lattice–driven CDW dynamics and their purely electronic counterparts.
\end{abstract}

\date{\today}

\maketitle

\section{Introduction}
\label{sec:intro}

Controlled quantum quenches---sudden changes in interaction strength, bandwidth, or external fields---have become a central tool for probing nonequilibrium dynamics in many-body quantum systems~\cite{Eisert_2015, Mitra_2018, Polkovnikov_2011}. Such protocols are realized in ultracold atomic gases confined in optical lattices as well as in solids driven by ultrafast pump-probe spectroscopy, where microscopic Hamiltonian parameters can be tuned on timescales shorter than the intrinsic dynamics of the system~\cite{Bloch2008, Cazalilla2011, Fausti_2011, Matsunaga2012, Matsunaga2013, mansart2013, matsunaga2014, mitrano2016}. These advances provide direct access to the real-time evolution of correlated quantum states far from equilibrium, enabling systematic investigations of post-quench unitary dynamics, thermalization in isolated systems, nonequilibrium transport, and dynamical phase transitions.

A central focus in this context is the quench dynamics of symmetry-breaking phases in the coherent regime. Following an interaction quench, the evolution of an ordered state is governed by the coupled dynamics of collective order parameters and quasiparticle excitations, which together determine the stability and temporal response of the broken-symmetry phase~\cite{Heyl_2014, Marino_2022}. Extensive theoretical work has explored such dynamics in interacting electron systems with long-range order, including BCS superconductors, charge- and spin-density-wave phases, and other complex orders~\cite{Barankov_2004,Barankov_2006,Foster_2013,Foster_2014,Blinov_2017,Tsuji_2013,Dong_2015,Peronaci_2015,Chou_2017,Yoon_2017,Hannibal_2018a,Hannibal_2018,Cui_2019,Scaramazza_2019,Chen_2020}. A key outcome of these studies is the identification of three generic dynamical regimes of the order parameter controlled by the quench amplitude: strong quenches lead to persistent, phase-coherent oscillations; intermediate quenches yield relaxation toward a finite nonequilibrium value with damped oscillations; and weak quenches result in overdamped dynamics with asymptotic decay of the order parameter.

While this classification is well established for purely electronic models with electron-electron interactions, many experimentally relevant systems involve additional slow degrees of freedom that can qualitatively modify the nonequilibrium response. In particular, electron-lattice coupling introduces retardation effects, extra channels for energy exchange, and additional intrinsic timescales. Charge-density-wave (CDW) systems provide a paradigmatic setting in which electronic ordering is intimately tied to lattice distortions~\cite{Gruner_1988, Thorne_1996}, raising the question of how coupled electronic and lattice dynamics reshape quench-induced behavior~\cite{Rohwer_2011, Hellmann_2012, Porer_2014, Hedayat_2019, Sayers_2020}. In this work, we address this issue by studying the interaction-quench dynamics of the CDW phase in the Holstein model~\cite{Scalettar_1989, Bonca_1999, Golez_2012, Mishchenko_2014, Costa_2018, Bradley_2021}, a minimal framework for electron-phonon driven charge order.

To make progress in a controlled and transparent setting, we adopt a semiclassical formulation of the Holstein model in which the lattice degrees of freedom are treated as classical variables. This approximation is motivated both by the difficulty of simulating fully quantum phonon dynamics far from equilibrium and by the observation that the essential features of CDW order in the Holstein model are already well captured in this limit~\cite{Esterlis_2019}. The resulting hybrid quantum-classical system is evolved in real time using Ehrenfest dynamics~\cite{Tully2023, Mitric2024}, where the electronic subsystem evolves quantum mechanically in the instantaneous lattice background, while the lattice coordinates follow classical equations of motion determined self-consistently by electronic expectation values. This coupled dynamics conserves the total energy and serves as a natural analog of unitary time evolution for hybrid quantum-classical systems.

Within this framework, we investigate the post-quench dynamics of checkerboard CDW order in the square-lattice Holstein model using the Anderson pseudospin representation~\cite{Anderson_1958}. It is worth noting that similar pseudospin formulation has been widely employed in the analysis of quench dynamics for symmetry-breaking phases of pure electronic systems~\cite{Barankov_2004,Barankov_2006,Foster_2013,Foster_2014,Blinov_2017}. Assuming spatial homogeneity and focusing on a single ordering wave vector $\mathbf{K}=(\pi,\pi)$, translational invariance allows the electronic sector to be mapped onto effective pseudospins in momentum space, where each pair of states $(\mathbf{p},\mathbf{p}+\mathbf{K})$ forms a two-level system described by a pseudospin vector $\mathbf{S}_{\mathbf{p}}$. The CDW order parameter and electronic dispersion define an effective pseudomagnetic field governing the precession of these pseudospins, reducing the semiclassical Holstein dynamics to a set of coupled precession equations. While real-space inhomogeneities can give rise to additional phenomena, this approach captures the intrinsic collisionless dynamics of CDW order and has been successfully applied to quench studies of superconductors and other symmetry-broken phases.

By tracking the coupled time evolution of the electronic CDW order parameter and the lattice displacement mode following an interaction quench, we identify three characteristic dynamical regimes analogous to those found in purely electronic systems. Notably, we find that persistent oscillations of the CDW order parameter remain robust due to coherent lattice motion. In addition, the explicit electron–phonon coupling induces a quench-dependent renormalization of the phonon frequency, illustrating how lattice dynamics qualitatively modify collisionless quench behavior in a $\mathbb{Z}_2$ symmetry-broken CDW system beyond the purely electronic limit. The homogeneous dynamics uncovered here provide a well-defined reference framework for post-quench CDW behavior, against which more intricate quench dynamics involving spatial inhomogeneity and pattern formation can be systematically compared in future studies.

The remainder of this paper is organized as follows. Section~\ref{sec:model} introduces the semiclassical Holstein model and derives the Anderson pseudospin formulation of the post-quench dynamics. Section~\ref{sec:result} presents the main numerical results for the electronic and lattice dynamics. In Sec.~\ref{sec:discussion}, we analyze the resulting dynamical regimes and discuss their physical interpretation and limitations. Section~\ref{sec:potential} constructs an effective nonequilibrium potential for the lattice mode and uses it to interpret the phonon dynamics. Finally, Sec.~\ref{sec:conclusion} summarizes our findings and outlines directions for future work.

\section{Pseudospin dynamics for the Holstein model}

\label{sec:model}

We consider the spinless Holstein model at half filling on a two-dimensional square lattice, described by the Hamiltonian
\begin{equation}
\hat{\mathcal{H}}=\hat{\mathcal{H}}_{\mathrm{e}} + \hat{\mathcal{H}}_{\mathrm{L}} + \hat{\mathcal{H}}_{\mathrm{eL}}.
\end{equation}
which consists of three contributions: the electronic kinetic energy, the lattice (phonon) degrees of freedom, and the electron–lattice coupling. The electron part, $\hat{\mathcal{H}}_{\mathrm{e}}$, describes the hopping of electrons between nearest-neighbor sites:
\begin{equation}
\hat{\mathcal{H}}_{\mathrm{e}}=-t_{\rm nn}\sum_{\langle ij\rangle}\left(\hat{c}^{\dag}_{i}\hat{c}_{j} + \hat{c}^{\dag}_{j}\hat{c}_{i} \right),
\end{equation}
where $c_i^\dagger$ ($c_i$) is the electron creation (annihilation) operator at site $i$ and its adjacent lattice sites $j$ with hopping constant $t_{\rm nn}$. The second term represents the lattice vibration energy, 
\begin{equation}
\hat{\mathcal{H}}_{\mathrm{L}}=\sum_{i}\left( \frac{1}{2m}\hat{P}^{2}_{i}+\frac{1}{2}m\Omega^2 \hat{Q}^{2}_{i} \right),
\end{equation}
where $\hat{Q}_i$ and $\hat{P}_i$ are the local lattice displacement and its conjugate momentum, respectively, with phonon mass $m$ and bare frequency $\Omega$. Finally, the electron–lattice coupling is given by
\begin{equation}
\hat{\mathcal{H}}_{\mathrm{eL}}=-g\sum_{i}\left(\hat{n}_{i}-\frac{1}{2}\right)\hat{Q}_{i}
\end{equation}
where $\hat{n}_i=\hat{c}_i^\dagger\hat{c}_i$ is the local fermion density and $g$ denotes the coupling strength. The coupling is written in a particle-hole symmetric form such that, at half filling, the equilibrium state satisfies $\langle \hat{n}_i\rangle = \tfrac{1}{2}$ and $\langle \hat{Q}_i\rangle = 0$ with vanishing chemical potential.

As discussed in Sec.~\ref{sec:intro}, previous numerical studies have shown that the semiclassical Holstein model captures the essential physics of charge-density-wave (CDW) order. This motivates us to adopt the semiclassical limit as a controlled framework for investigating post-quench dynamics in the collisionless regime. Within this approach, we describe the time-dependent quantum state of the coupled electron–phonon system by a product ansatz,
\begin{equation}
\ket{\Gamma(t)} = \ket{\Phi(t)} \otimes \ket{\Psi(t)},
\end{equation}
where $\ket{\Phi(t)}$ and $\ket{\Psi(t)}$ denote the phonon and electronic many-body wave functions, respectively.

We further treat the phonon sector at the mean-field level, assuming that the phonon wave function factorizes over lattice sites, $\ket{\Phi(t)}=\prod_{i} \ket{\phi_{i}(t)}$. Within this approximation, the lattice displacement and momentum at site $i$ are characterized by their expectation values, $Q_{i}=\bra{\phi_{i}} \hat{Q}_{i} \ket{\phi_{i}}$ and $P_{i}=\bra{\phi_{i}} \hat{P}_{i} \ket{\phi_{i}}$. The equations of motion for these classical variables follow from the Heisenberg equations evaluated with respect to the full product state $\ket{\Gamma(t)}$. Equivalently, they can be obtained from the Ehrenfest theorem, which governs the coupled evolution of classical observables and quantum expectation values. This procedure yields
\begin{eqnarray}
	\label{eq:newton}
	\frac{dQ_i}{dt} = \frac{P_i}{m}, \qquad \frac{dP_i}{dt} = g \left(\langle n_i\rangle - \frac{1}{2} \right) - m\Omega^2 Q_i,
\end{eqnarray}
which are the Newtonian equations for the classical lattice degrees of freedom.

In the interaction-quench protocol considered in this work, the initial electronic many-body state within the semiclassical approximation is a Slater determinant. Because the Holstein Hamiltonian is bilinear in the fermionic creation and annihilation operators, an initial Slater determinant remains a Slater determinant throughout the subsequent time evolution governed by this quadratic Hamiltonian~\cite{Luo_2021}.
While this property allows, in principle, for an exact description of the electronic dynamics by evolving the full time-dependent Slater determinant~\cite{Koshibae_2011,Ono_2017}, a more efficient approach is to formulate the dynamics in terms of the single-particle density matrix,
\begin{eqnarray}
	\rho_{ij}(t) = \bra{\Psi(t)} \hat{c}^\dagger_j \hat{c}^{\,}_{i} \ket{\Psi(t)}.
\end{eqnarray}
which fully characterizes the many-body state in the absence of interactions beyond quadratic order.
The time evolution of $\rho(t)$ is governed by the von Neumann equation,
\begin{equation}
    \frac{d\rho}{dt} = \frac{-i}{\hbar} [H(\{Q_i\}),\rho],
\end{equation}
where $H(\{Q_i\})$ is the first-quantized Hamiltonian 
\begin{eqnarray}
	H_{ij} = -t_{ij} - g Q_i \delta_{ij},
\end{eqnarray}
Applying the von Neumann equation to the single-particle density matrix yields the explicit equation of motion for the electronic degrees of freedom,
\begin{eqnarray}
\label{eq:von-neumann}
    i\hbar \frac{d\rho_{ij}}{dt} =  \sum_{k}\left( \rho_{ik}t_{kj} - t_{ik}\rho_{kj} \right) + g\left( Q_{j} - Q_{i} \right)\rho_{ij}. \quad
\end{eqnarray}
which governs the coupled evolution of electrons in the presence of a time-dependent lattice background.

Under the assumption that the system remains spatially homogeneous following the interaction quench—thereby preserving translational invariance—the electronic dynamics can be formulated most naturally in momentum space. In this regime, spatial inhomogeneities and domain formation are explicitly excluded, and the electronic state is fully characterized by momentum-resolved quantities.
Performing a Fourier transformation of the fermionic operators, $\hat{c}_{i} = \frac{1}{\sqrt{N}} \sum_k \hat{c}_{\mathbf k} \exp(i\mathbf{k}\cdot \mathbf{r}_i)$, and expressing the von Neumann equation in the momentum basis yields the equation of motion for the single-particle density matrix,
\begin{equation}
\label{von_Neumann_pq}
\begin{aligned}
\frac{d\rho_{\mathbf{p},\mathbf{q}}}{dt}=\frac{i}{\hbar}&\left[\left(\epsilon_{\mathbf{q}}-\epsilon_{\mathbf{p}} \right)\rho_{\mathbf{p},\mathbf{q}}\right. \\
&+\left. gQ_{\mathbf{K}}\left( \rho_{\mathbf{p}-\mathbf{K}, \mathbf{q}}-\rho_{\mathbf{p},\mathbf{q}+\mathbf{K}} \right)\right]
\end{aligned}
\end{equation}
where $\epsilon_{\mathbf{q}} = -2 t_{\rm nn} \left[ \cos(q_x) +\cos(q_y) \right]$, is the dispersion relation of noninteracting electrons on a square lattice. In writing Eq.~\eqref{von_Neumann_pq}, we have assumed that the lattice distortion is dominated by a single ordering wave vector $\mathbf{K}=(\pi,\pi)$, reflecting the leading charge-density-wave (CDW) instability of the Holstein model at half filling.

This momentum-space structure naturally suggests a reformulation of the dynamics in terms of an effective pseudospin representation. Following Anderson’s pseudospin construction, originally introduced to describe pairing dynamics in superconductors, we reinterpret the CDW instability as a coupling between electronic states at momenta $\mathbf{p}$ and $\mathbf{p}+\mathbf{K}$. In the present context, this pairing reflects charge modulation between the two sublattices of the square lattice and does not involve the spin degree of freedom. Specifically, we introduce pseudospin operators in the reduced Brillouin zone,
\begin{eqnarray}
	& & \hat S_{\mathbf{p}}^{+} = \hat{c}_{\mathbf{p}+\mathbf{K}}^{\dagger} c^{\,}_{\mathbf{p}}, \quad \hat S_{\mathbf{p}}^{-} = \hat{c}_{\mathbf{p}}^{\dagger} \hat{c}^{\,}_{\mathbf{p}+\mathbf{K}}, \nonumber \\
	& & \hat S_{\mathbf{p}}^{z} = (\hat n_{\mathbf{p}} - \hat n_{\mathbf{p}+\mathbf{K}})/2,
\end{eqnarray}
which act on the two-level subspace spanned by ${ \ket{\mathbf{p}}, \ket{\mathbf{p}+\mathbf{K}} }$. These operators obey the standard SU(2) algebra, $[\hat S_{\mathbf{p}}^{+},\hat S_{\mathbf{p}}^{-}] = 2 \hat S_{\mathbf{p}}^{z}$.
and provide a compact and physically transparent framework for describing the coherent post-quench dynamics of the homogeneous CDW order parameter.
Within this formulation, the expectation values of the pseudospin components are directly expressed in terms of the single-particle density matrix,
\begin{eqnarray}
	& & S_{\mathbf{p}}^{x} = \mathrm{Re}\,\rho_{\mathbf{p},\mathbf{p}+\mathbf{K}},\quad S_{\mathbf{p}}^{y} = \mathrm{Im}\,\rho_{\mathbf{p},\mathbf{p}+\mathbf{K}}, \nonumber \\
	& & S_{\mathbf{p}}^{z} = (\rho_{\mathbf{p},\mathbf{p}} - \rho_{\mathbf{p}+\mathbf{K},\mathbf{p}+\mathbf{K}})/2.
\end{eqnarray}
making explicit the equivalence between the pseudospin dynamics and the evolution of the momentum-space density matrix.
By recasting Eq.~(\ref{von_Neumann_pq}) in terms of these pseudospin variables, we obtain the following set of coupled equations of motion:
\begin{subequations}
\label{eq:pseudospin}
\begin{align}
\frac{dS^{z}_\mathbf{p}}{dt} &= \frac{2g}{\hbar}\,Q_{\mathbf{K}}\,S^{y}_\mathbf{p}, \label{eq:pseudospin_a} \\
\frac{dS^{x}_\mathbf{p}}{dt} &= \frac{2}{\hbar}\,\epsilon_\mathbf{p}\,S^{y}_\mathbf{p}, \label{eq:pseudospin_b} \\
\frac{dS^{y}_\mathbf{p}}{dt} &= -\frac{2}{\hbar}\,\epsilon_\mathbf{p}\,S^{x}_\mathbf{p} - \frac{2}{\hbar}g\,Q_{\mathbf{K}}\,S^{z}_\mathbf{p}. \label{eq:pseudospin_c}
\end{align}
\end{subequations}
which can be written compactly in the Bloch form,
\begin{equation}
\label{eq:Bloch_eq}
\frac{d\mathbf{S}_\mathbf{p}}{dt}=\frac{1}{\hbar}\mathbf{B}_\mathbf{p}\times\mathbf{S}_\mathbf{p}
\end{equation}
with an effective pseudomagnetic field 
\begin{eqnarray}
	\mathbf{B}_\mathbf{p}=(2gQ_{\mathbf{K}}, 0, -2\epsilon_\mathbf{p}). 
\end{eqnarray}
Owing to translational symmetry, the $N^2$ degrees of freedom of the single-particle density matrix reduce to $3N$ variables, corresponding to one three-component pseudospin $\mathbf{S}_{\mathbf{p}}$ for each momentum $\mathbf{p}$.

The lattice dynamics in reciprocal space follow analogously from Eq.~(\ref{eq:newton}) upon Fourier transformation,
\begin{eqnarray}
	\label{eq:newton_eq}
	\frac{dQ_{\mathbf{K}}}{dt} = \frac{P_{\mathbf{K}}}{m}, \qquad \frac{dP_{\mathbf{K}}}{dt} = g \Delta_{\mathbf K} - m\Omega^2 Q_{\mathbf{K}},
\end{eqnarray}
where the checkerboard CDW order $\Delta_{\mathbf K}$ defined as 
\begin{equation}
\label{eq:CDW}
\Delta_{\mathbf{K}}=\frac{1}{N} \sum_{\mathbf{p}}\rho_{\mathbf{p}, \mathbf{p}+\mathbf{K}}.
\end{equation}
acts as the electronic driving force on the collective harmonic oscillator describing the staggered lattice distortion $Q_{\mathbf{K}}$. Taken together, Eqs.~(\ref{eq:Bloch_eq}) and~(\ref{eq:newton_eq}) constitute a closed set of equations governing the coupled, collisionless dynamics of the electronic pseudospins and the lattice distortion following the interaction quench.

The nonequilibrium dynamics of the semiclassical  Holstein model are governed by two intrinsic timescales associated with the electronic and lattice subsystems. Electronic motion is controlled by the nearest-neighbor hopping amplitude $t_{\rm nn}$, which defines an electronic timescale $\tau_{\rm e} = \hbar / t_{\rm nn}$ and an associated bandwidth $W = 8 t_{\rm nn}$. This timescale sets the characteristic rate at which electrons propagate through the lattice. In contrast, lattice dynamics are determined by the Einstein phonon frequency $\Omega$, giving rise to a lattice timescale $\tau_{\rm L} = 1 / \Omega$. The relative separation of these timescales is quantified by the dimensionless adiabatic parameter
$r = \tau_{\rm e} / \tau_{\rm L} = \hbar \Omega / t_{\rm nn}$. Small values of $r$ correspond to the adiabatic regime, in which lattice dynamics are slow compared to electronic motion, while larger values indicate increasingly nonadiabatic behavior.

In addition to these timescales, it is useful to introduce characteristic scales for the lattice sector. A representative lattice distortion amplitude $Q^*$ can be estimated by comparing the elastic cost $KQ^{2}$ with the electron–phonon interaction energy $g\,\langle n \rangle Q$. Since $\langle n \rangle$ simply sets a dimensionless density scale of order unity, one obtains the characteristic amplitude $Q^* \equiv g / K$. The corresponding momentum scale follows from the Newton equations of motion in Eq.~(\ref{eq:newton_eq}), giving $P^* = m \Omega Q^*$. The strength of the electron–phonon interaction is characterized by the dimensionless coupling constant $\lambda = g Q^* / W = g^2 / (W K)$, which measures the characteristic electron–phonon energy relative to the electronic bandwidth. In the simulations presented here, we fix the adiabatic parameter to $r = 0.3$ and vary the coupling strength over the range $\lambda \in [0, 0.2]$ to explore different dynamical regimes. Unless otherwise stated, all results are reported in natural units: time is measured in units of $\tau_{\rm e}$, energies in units of $t_{\rm nn}$, and lattice displacements in units of $Q^*$. This choice of dimensionless variables facilitates a transparent comparison of the relative roles of electronic and lattice dynamics in the nonequilibrium evolution of the system. Further details on the dimensionless formulation of the coupled dynamics are provided in Appendix~\ref{app:dimensional}.

\begin{figure*}[t!]
\includegraphics[width=180mm]{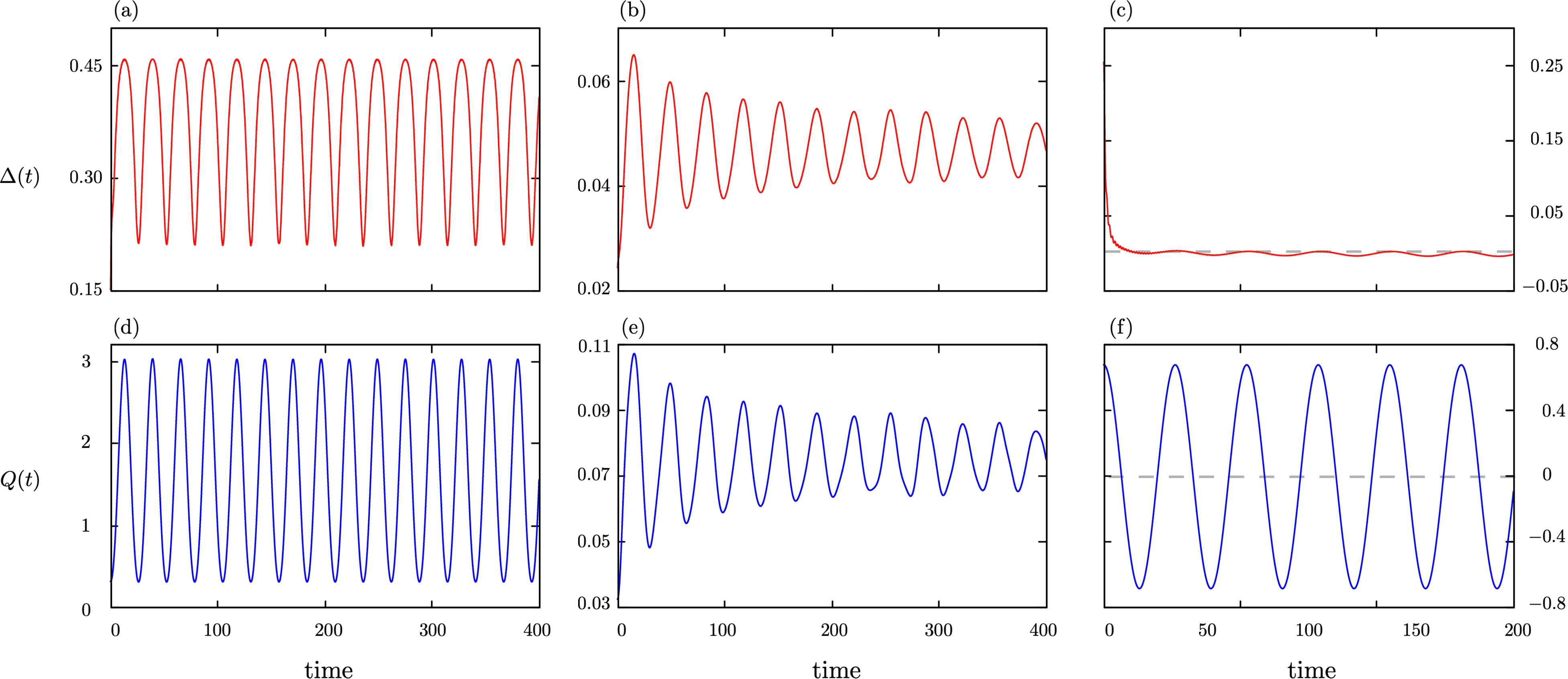}
\caption{\label{dynamical_regime} Time evolution of the CDW order parameter $\Delta_{\mathbf{K}}(t)$ (top panels) and the staggered lattice distortion $Q_{\mathbf{K}}(t)$ (bottom panels) following a sudden quench of the electron–phonon coupling. Representative results are shown for the three distinct dynamical regimes: phase-locked oscillations (left), Landau-damped dynamics (middle), and nominally overdamped dynamics (right). The electronic CDW order parameter $\Delta_{\mathbf{K}}$ is dimensionless, while the lattice distortion $Q_{\mathbf{K}}$ is measured in units of the characteristic scale $Q^{*} = g/k$. Time is expressed in units of $\tau_e = \hbar / t_{\rm nn}$. In terms of the dimensionless electron–phonon coupling $\lambda = g^{2}/(W K)$, the corresponding quench parameters are $\lambda_i = 0.08 \rightarrow \lambda_f = 0.32$ (phase-locked), $\lambda_i = 0.03 \rightarrow \lambda_f = 0.045$ (Landau-damped), and $\lambda_i = 0.125 \rightarrow \lambda_f = 0.0013$ (nominally overdamped).}
\end{figure*}

\section{Post-quench order parameter dynamics}

\label{sec:result}

In this section, we analyze the post-quench dynamics of the semiclassical Holstein model within the Anderson pseudospin framework. The system is initially prepared in the CDW ground state corresponding to an electron–phonon coupling $g_i$. In this initial state, the pseudspins  $\mathbf{S}_{\mathbf{p}}$ are aligned with the effective magnetic field $\mathbf{B}_{\mathbf{p}}=(2g_i Q_{\mathbf{K}}^{(i)},0,\epsilon_\mathbf{p})$, where $Q_{\mathbf{K}}^{(i)}$ denotes the initial staggered lattice distortion. At time $t=0$, the coupling is suddenly quenched to a new value $g_f$, and for $t>0$ the coupled lattice and pseudospin dynamics are evolved with the new electron-lattice coupling $g_f$ according to their respective equations of motion. We follow the subsequent real-time evolution of the CDW order parameter $\Delta_{\mathbf{K}}(t)$ and the lattice distortion $Q_{\mathbf{K}}(t)$.

Since the initial state is not the ground state of the post-quench Hamiltonian with $g_f$, the quench injects a finite amount of energy into the otherwise closed system. Assuming that spatially homogeneous CDW order remains dominant during the evolution, this excess energy drives a nonlinear, self-consistent dynamics between the lattice displacement and the pseudospin degrees of freedom. Depending on the quench parameters $(g_i, g_f)$, this coupled dynamics gives rise to qualitatively distinct temporal behaviors of the CDW order parameter.

Figure~\ref{dynamical_regime} summarizes the post-quench dynamics of the CDW order parameter and the associated lattice distortion within the pseudospin formulation. The upper panels show the time evolution of the electronic CDW order parameter $\Delta_{\mathbf{K}}(t)$, while the lower panels display the corresponding staggered lattice displacement $Q_{\mathbf{K}}(t)$. Depending on the relative values of the initial and final electron–phonon couplings $(g_i, g_f)$, we identify three qualitatively distinct dynamical regimes. These regimes closely parallel the collisionless quench dynamics of BCS superconductors studied in Ref.~\cite{Barankov_2006}, with an important distinction arising from the presence of dynamical lattice degrees of freedom.

For sufficiently strong quenches with $g_f \gg g_i$, the system enters a phase-locked oscillatory regime, shown in Figs.~\ref{dynamical_regime}(a) and (d). In this regime, both $\Delta_{\mathbf{K}}(t)$ and $Q_{\mathbf{K}}(t)$ exhibit persistent, nondecaying oscillations with well-defined amplitudes and frequencies. The electronic pseudospins remain synchronized and precess coherently around the effective pseudomagnetic field, producing bounded oscillations of the CDW order parameter. The lattice displacement follows this motion in a phase-locked manner, reflecting strong feedback between electronic ordering and lattice distortion. This behavior is directly analogous to the synchronized, undamped oscillations of the pairing amplitude in BCS quench dynamics and corresponds to the collisionless, nonthermal regime of the coupled electron–phonon system.

At intermediate quench strengths, where $g_f$ exceeds but is comparable to $g_i$, the system crosses over into a Landau-damped regime, illustrated in Figs.~\ref{dynamical_regime}(b) and (e). Here, $\Delta_{\mathbf{K}}(t)$ displays oscillations with a decaying envelope, approaching a finite nonequilibrium steady-state value. As in the BCS case, this damping originates from dephasing among the pseudospins due to their dispersion in precession frequencies. However, in contrast to purely electronic systems, the oscillations in both $\Delta_{\mathbf{K}}(t)$ and $Q_{\mathbf{K}}(t)$ do not disappear entirely at long times. Instead, the presence of a coherent lattice mode sustains residual oscillations indefinitely, preventing complete decay and stabilizing a persistently oscillatory nonequilibrium state.

For sufficiently weak quenches $g_f \ll g_i$, the dynamics enters a regime that is nominally overdamped, as shown in Figs.~\ref{dynamical_regime}(c) and (f). In this case, the CDW order parameter rapidly decays toward zero, indicating the loss of long-range electronic CDW order due to strong pseudospin dephasing. Nevertheless, unlike the overdamped regime in BCS quench dynamics—where the order parameter relaxes monotonically without oscillations—the coupled electron–phonon system continues to exhibit sustained oscillatory behavior. The lattice displacement undergoes large-amplitude oscillations about zero, which in turn induce weak but persistent oscillations in the electronic sector. Thus, even in the regime where electronic order is dynamically destroyed on average, the lattice dynamics ensures that the system never fully relaxes to a static state.

\begin{figure*}[t!]
\includegraphics[width=180mm]{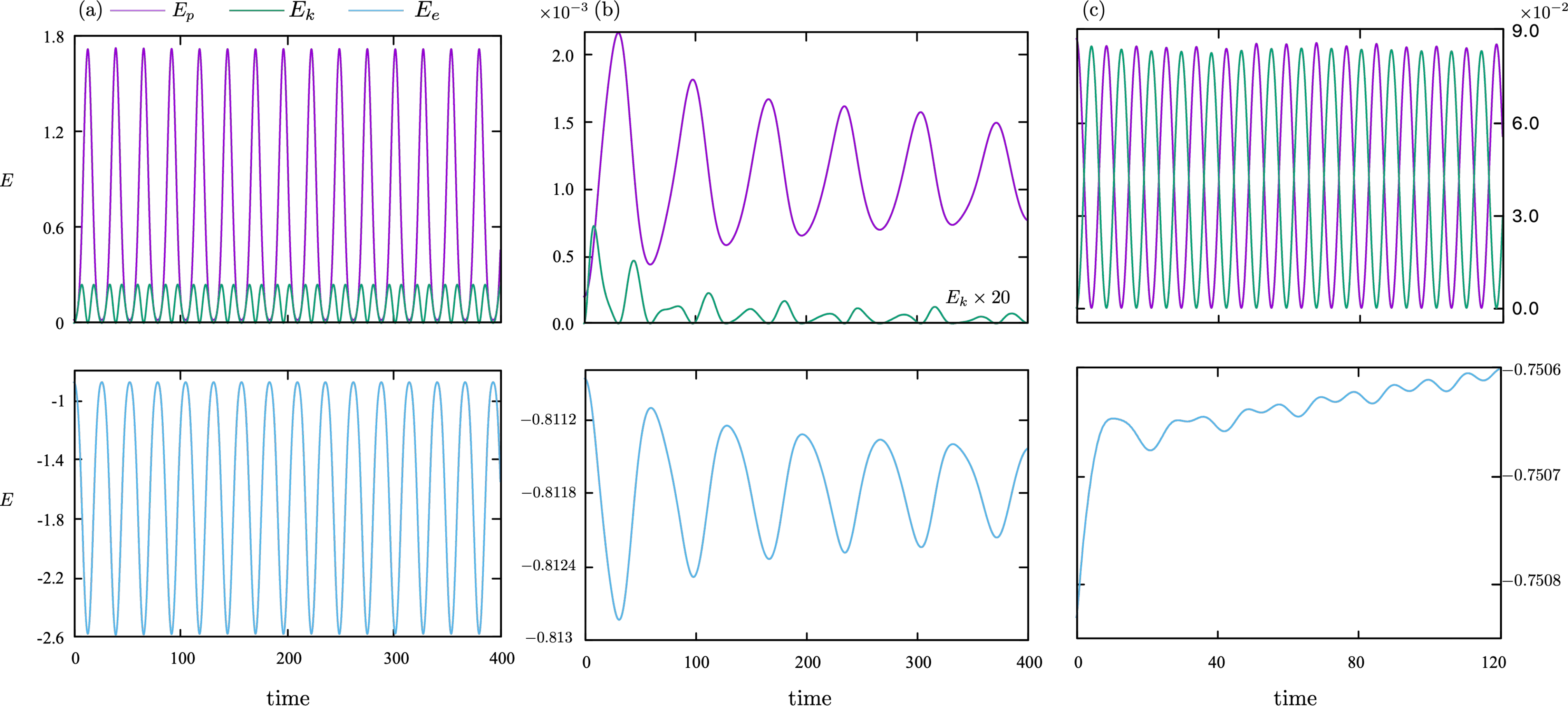}
\caption{\label{fig:energy_dynamics} The time dependence of the potential energy $E_p$ (purple curves), kinetic energy $E_k$ (green curves), and electronic energy $E_e$ (blue curves) for the three dynamical regimes shown in Fig.~\ref{dynamical_regime}. The potential energy $E_p$ is measured in units of $\frac{1}{2}(Q^{*})^2$, the lattice kinetic energy $E_k$ in units of $\frac{1}{2}(P^*)^2$, and the electronic energy in units of $t_{\rm nn}$. }
\end{figure*}

To further elucidate the post-quench dynamics of the electronic and lattice subsystems, we examine the time evolution of the electronic energy $E_e$ and the lattice kinetic and potential energies, $E_k$ and $E_p$, respectively. These quantities are defined as
\begin{eqnarray}
 \label{eq:energies}
    \begin{aligned}
    &E_e = \frac{1}{N} \sum_{\mathbf{p}} \left( \epsilon_{\mathbf{p}} \, \rho_{\mathbf{p},\mathbf{p}} - g \, Q_{\mathbf{K}}\, \rho^{\,}_{\mathbf{p}, \mathbf{p+K}} \right),\\
    &E_k = \frac{1}{2m}P^{2}_{\mathbf{K}},\qquad
    E_p = \frac{1}{2}m\Omega^2 Q^{2}_{\mathbf{K}},   
    \end{aligned}
\end{eqnarray}
The time evolution of the three energy components are shown in Fig.~\ref{fig:energy_dynamics} for the three post-quench dynamical regimes identified in Fig.~\ref{dynamical_regime}. The left, middle, and right panels correspond to the phase-locked oscillatory, Landau-damped, and overdamped regimes, respectively. While the total energy is conserved in all cases, the figure highlights how energy is dynamically redistributed between electronic and lattice degrees of freedom.

In the phase-locked regime [Fig.~\ref{fig:energy_dynamics}(a)], all three energy components exhibit persistent oscillations with essentially constant amplitude, reflecting a fully synchronized evolution of the pseudospin and lattice sectors. The absence of decay indicates a strongly nonthermal state, in which the dynamics follow a stable, quasi-integrable orbit in phase space rather than relaxing toward ergodicity. This behavior closely parallels the undamped Rabi-like oscillations characteristic of collisionless BCS quench dynamics.

In the Landau-damped regime shown in Fig.~\ref{fig:energy_dynamics}(b), the decay of the order-parameter oscillations [Fig.~~\ref{dynamical_regime}(b,e)] is mirrored in the energy sector by a gradual suppression of the lattice energies $E_p$ and $E_k$. Pseudospin dephasing transfers energy from the coherent lattice-CDW mode into a broad continuum of incoherent electronic excitations.
The over-damped regime in Fig.~\ref{fig:energy_dynamics}(c) most clearly highlights the emergence of coherent phonons after the collapse of electronic CDW order. Following the strong quench, rapid pseudospin dephasing suppresses the order parameter $\Delta_K(t)$, and the electronic energy $E_e$ correspondingly relaxes to a nearly time-independent plateau, indicating a dephased, prethermal electronic state. In contrast, the lattice sector retains a weak but clearly discernible harmonic oscillation, consistent with the persistent lattice dynamics shown in Fig.~\ref{dynamical_regime}(f).

The persistence of oscillations driven by lattice degrees of freedom is most transparently illustrated in the overdamped regime with $g_f \ll g_i$. In this limit, $g_f$ can be treated as a small parameter, and the coupled Newton and Bloch equations may be solved perturbatively. The zeroth-order (unperturbed) solution corresponds to the limit $g_f = 0$, in which the electron–phonon coupling is suddenly switched off. In this case, the lattice retains a finite displacement $Q_{\mathbf K}^{(i)}$ at $t=0$ and subsequently undergoes persistent harmonic motion. The solution for the lattice dynamics can then be written as
\begin{eqnarray}
	Q_{\mathbf K}(t) =   Q_{\mathbf K}^{(i)} \cos(\Omega t) + \delta Q_{\mathbf K}(t),
\end{eqnarray}
where $\delta Q_{\mathbf K}(t)$ denotes the first-order correction arising from a small but finite post-quench coupling $g_f$. Similarly, the CDW order parameter admits the perturbative expansion
\begin{eqnarray}
	\Delta_{\mathbf K}(t) = \Delta^{(0)}_{\mathbf K}(t) + \delta\Delta_{\mathbf K}(t), 
\end{eqnarray}
The zeroth-order solution $\Delta^{(0)}_{\mathbf K}(t)$ describes the dephasing of pseudospins that were initially aligned by the CDW order associated with $Q_{\mathbf K}^{(i)}$. As shown in Appendix~\ref{app:zero-gf}, this dephasing leads to a power-law decay of the CDW order,
\begin{eqnarray}
	\Delta^{(0)}_{\mathbf K}(t) = \sum_{\mathbf p} S^x_{\mathbf p}(0) e^{2 i \epsilon_{\mathbf p} t} \sim \frac{1}{t}.
\end{eqnarray}
reflecting the absence of any restoring electronic coherence in the uncoupled limit.
We next consider the first-order corrections. The correction to the lattice dynamics $\delta Q_{\mathbf K}(t)$ is obtained by treating the zeroth-order CDW order $\Delta^{(0)}_{\mathbf K}(t)$ as a source term in the equation of motion of a simple harmonic oscillator, yielding
\begin{eqnarray}
	\delta Q_{\mathbf K}(t) = \frac{1}{\Omega} \int_0^t \sin[\Omega(t-t')]\, \Delta^{(0)}_{\mathbf K}(t') dt'
\end{eqnarray}
At long times, this integral produces an oscillatory response, $\delta Q_{\mathbf K}(t) \sim \cos(\omega t + \phi_0)$ with an frequency $\omega \sim \Omega$, indicating that the lattice sustains coherent oscillations even in the presence of electronic dephasing. The corresponding first-order correction to the CDW order parameter follows from the electronic susceptibility,
\begin{eqnarray}
	\delta \Delta_{\mathbf K}(t) = g_f Q_{\mathbf K}^{(i)} \int_0^t \chi(t-t') \cos(\Omega t') dt'
\end{eqnarray}
where the electron response function is given by
\begin{eqnarray}
	\chi(t) = \sum_{\mathbf p} \left( \rho_{\mathbf p, \mathbf p} - \rho_{\mathbf p + \mathbf K, \mathbf p + \mathbf K} \right) e^{i 2 \epsilon_{\mathbf p} t}.
\end{eqnarray}
This leads similarly to an oscillatory time dependence with a frequency close to $\Omega$. Importantly, although the zeroth-order CDW order $\Delta^{(0)}_{\mathbf K}(t)$ vanishes asymptotically as $1/t$, the feedback from the lattice motion encoded in the first-order correction $\delta \Delta_{\mathbf K}(t)$ generates persistent oscillations of the CDW order at long times. While the present perturbative analysis is strictly valid only in the limit $g_f \ll g_i$, it strongly suggests that similar lattice-induced persistent oscillations should also be present in the Landau-damped regime, where $g_f \lesssim g_i$.

\section{Renormalized Lattice dynamics} 
\label{sec:discussion}

In contrast to quantum quenches in purely electronic systems—such as the well-studied BCS quench dynamics—the presence of lattice degrees of freedom introduces an intrinsic time scale $\tau_{\rm L} = 1/\Omega$, set by the bare oscillation frequency $\Omega$ of the Holstein phonons. The persistent oscillations discussed in the previous section are therefore naturally tied to this lattice time scale. At the same time, the coupling between electrons and lattice degrees of freedom, or equivalently between pseudospins and phonons, is expected to renormalize the effective oscillation frequency and modify the long-time dynamics. In this section, we present a systematic study of this electron-induced renormalization of the oscillation frequency following a quench.

To this end, we analyze post-quench dynamics for different values of the bare phonon frequency $\Omega$, while keeping the quench protocol fixed at $\lambda_i = 0.00125 \rightarrow 0.03$. For each value of $\Omega$, this procedure determines the corresponding equilibrium lattice distortions $Q^{(i)}_{\mathbf{K}}$ and $Q^{(f)}_{\mathbf{K}}$. As discussed in Sec.~\ref{sec:model}, the dynamics of the Holstein model are governed by two dimensionless parameters: the electron–phonon coupling $\lambda$ and the adiabaticity ratio $r = \tau_{\rm e}/\tau_{\rm L} = \hbar \Omega / t_{\rm nn}$. Varying the phonon frequency thus effectively tunes the adiabaticity of the system. However, to emphasize the physical origin of the frequency renormalization and facilitate direct comparison with the bare lattice mode, we present our results in terms of $\Omega$ itself. Accordingly, all frequencies are expressed in units of $t_{\rm nn}/\hbar$.

\begin{figure}[t]
\includegraphics[width=0.8\columnwidth]{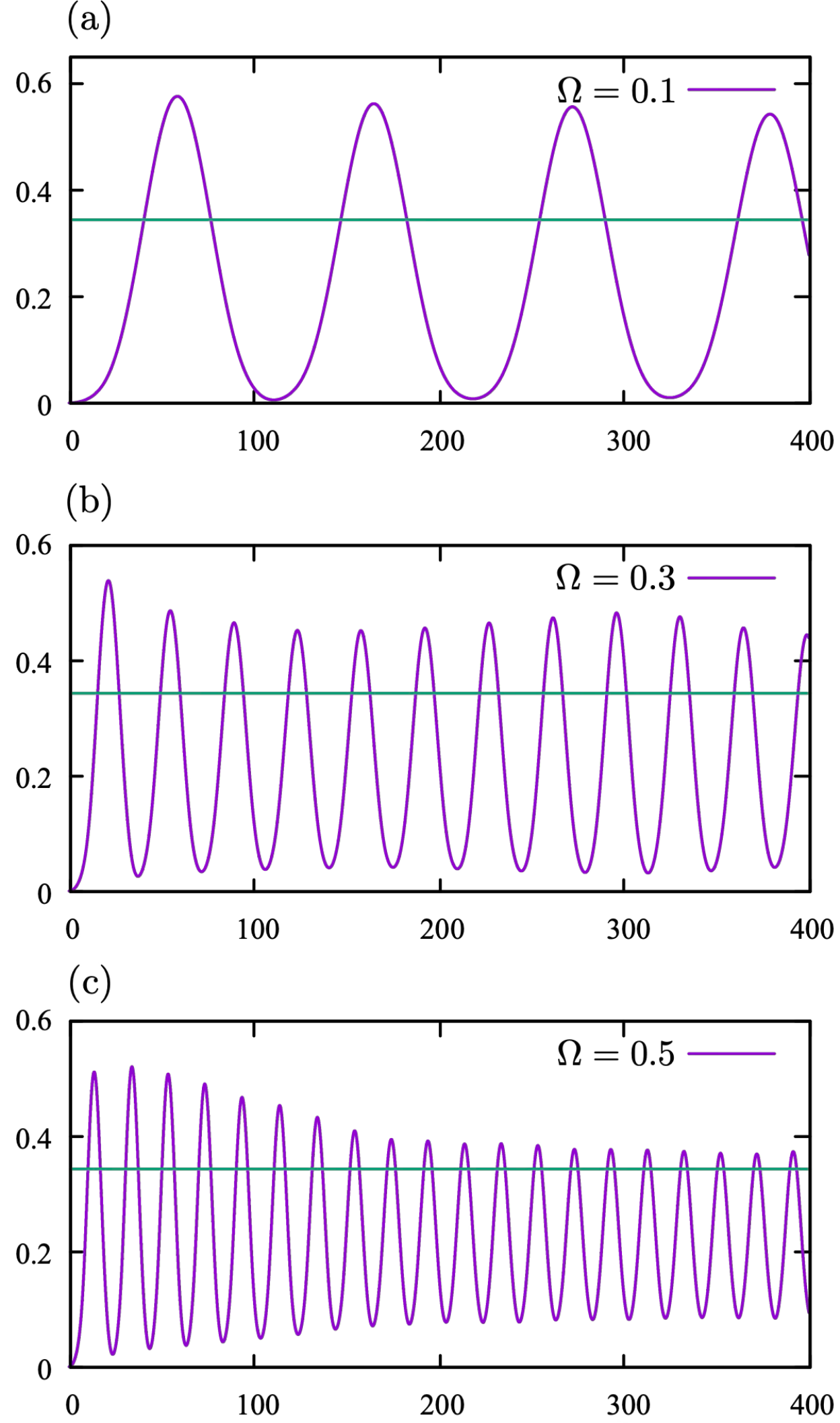}
\caption{\label{eta} Post-quench dynamics of the staggered lattice distortion $Q_{\mathbf{K}}(t)$ following an interaction quench $\lambda_i = 0.00125 \rightarrow 0.03$. The corresponding equilibrium lattice distortions are $Q^{(i)}_{\mathbf{K}} \simeq 0.06$ and $Q^{(f)}_{\mathbf{K}} \simeq 1.94$, measured in units of $Q^{*} = g/K$. Results are shown for several bare phonon frequencies of the classical lattice mode: (a) $\Omega = 0.1$, (b) $\Omega = 0.5$, and (c) $\Omega = 0.8$. The post-quench equilibrium value $Q^{(f)}_{\mathbf{K}}$ is indicated by the horizontal green line.}
\end{figure}

Fig.~\ref{eta} shows the post-quench evolution of the lattice distortion $Q(t)$ at three different values of $\Omega$. For small phonon frequency, exemplified by $\Omega = 0.1$ in Fig.~\ref{eta}(a), the lattice distortion exhibits long-lived oscillations with nearly constant amplitude, characteristic of a weakly damped, phase-coherent regime. As $\Omega$ is increased, however, the oscillations progressively lose amplitude and relax toward a smaller but finite value at long times, as seen in Figs.~\ref{eta}(b) and \ref{eta}(c). This trend indicates that faster lattice dynamics enhance the effective damping of the coupled CDW–phonon mode, consistent with stronger energy transfer from the lattice into incoherent electronic excitations.

\begin{figure}[t]
\includegraphics[width=0.8\columnwidth]{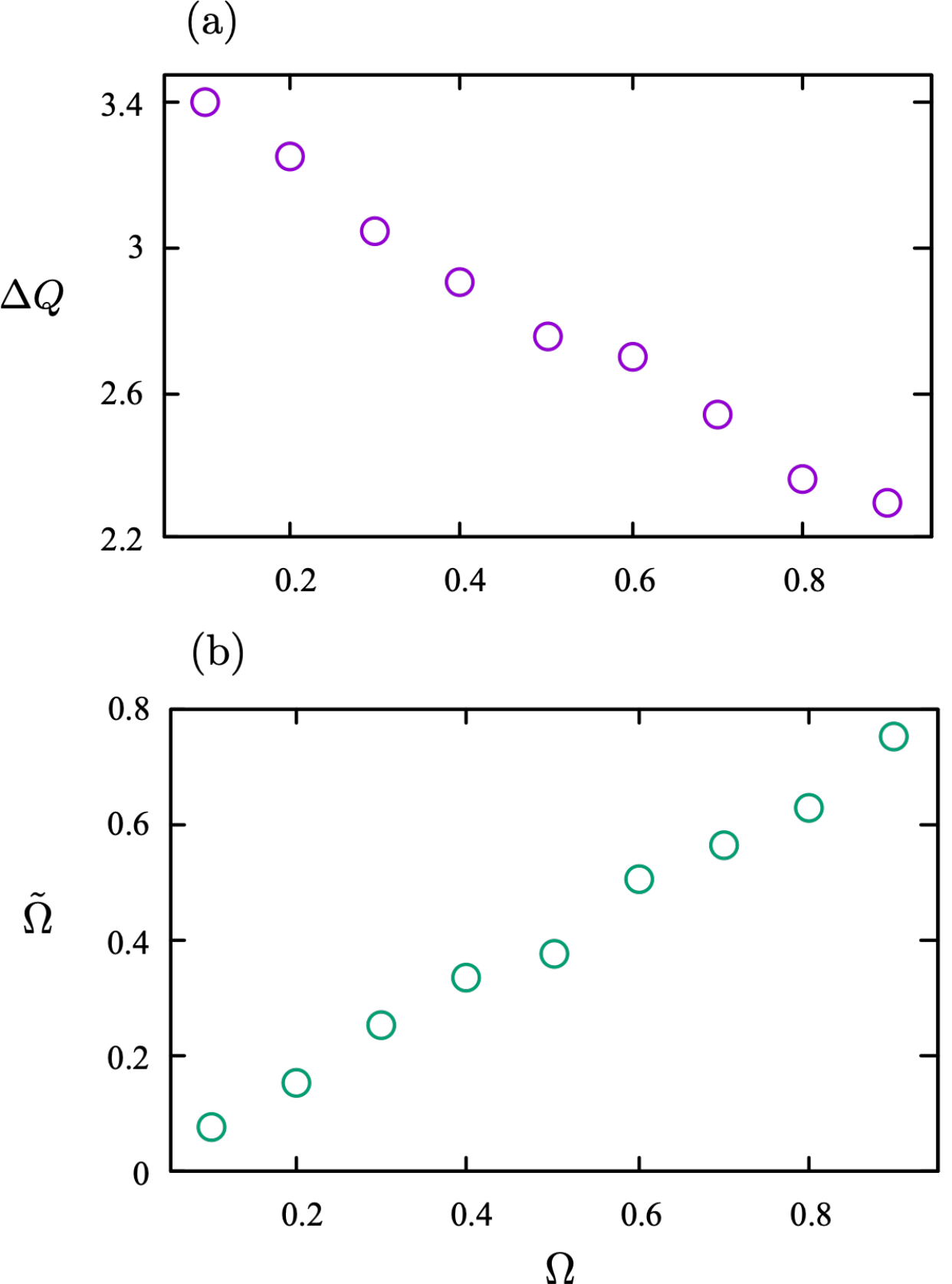}
\caption{\label{K0} Post-quench lattice dynamics as a function of the bare phonon frequency $\Omega$. (a) Oscillation amplitude $\Delta Q$ and (b) renormalized oscillation frequency $\tilde{\Omega}$ extracted from fitting of $Q_{\mathbf K}(t)$ from pseudo-spin simulations. Increasing $\Omega$ suppresses the oscillation amplitude and leads to a systematic reduction $\tilde{\Omega} < \Omega$ due to electron–phonon coupling.}
\end{figure}

To quantify this behavior, we extract the oscillation amplitude $\Delta Q$ of the lattice distortion for each value of $\Omega$. The resulting dependence $\Delta Q(\Omega)$ is shown in Fig.~\ref{K0}(a). A clear monotonic decrease of $\Delta Q$ with increasing $\Omega$ is observed. This trend indicates that slower lattice modes (small $\Omega$) support large-amplitude, weakly damped oscillations, whereas faster lattice dynamics lead to a substantial suppression of the oscillation amplitude. Physically, increasing $\Omega$ enhances the efficiency of energy transfer from the coherent lattice motion into the electronic degrees of freedom, thereby reducing the ability of the lattice to sustain large-amplitude oscillations after the quench.

Fig.~\ref{K0}(b) shows the renormalized oscillation frequency $\tilde{\Omega}$ extracted numerically from the time dependence of $Q_{\mathbf K}(t)$, plotted as a function of the bare phonon frequency $\Omega$. While $\tilde{\Omega}$ increases monotonically with $\Omega$, it remains systematically smaller than the bare value over the entire range shown. This deviation directly reflects the dynamical renormalization of the lattice mode induced by its coupling to the electronic subsystem. Even though the lattice is treated classically, the electronic backaction effectively softens the phonon dynamics, leading to a reduced oscillation frequency in the nonequilibrium steady state.

A simple, intuitive interpretation of the frequency renormalization can be obtained by examining the Newton equation of motion Eq.~(\ref{eq:newton}) for the staggered lattice mode. In the presence of electron-phonon coupling, the equation of motion may be written as
\begin{eqnarray}
	& & \frac{d^2 Q_{\mathbf k}}{dt^2} = -\Omega^2 Q_{\mathbf K} + \frac{g}{m} \Delta_{\mathbf K} \nonumber \\
	& & \qquad  = -\left( \Omega^2 - \frac{g}{m} \frac{\partial \Delta_{\mathbf K}}{\partial Q_{\mathbf K}} \right) Q_{\mathbf K}, \quad
\end{eqnarray}
In obtaining the second equality, we have assumed the electronic CDW order $\Delta_{\mathbf K}$ as a functional of the lattice displacement $Q_{\mathbf K}$, and expanded  it to linear order in $Q_{\mathbf K}$. Within this approximation, the coupling to the electronic degrees of freedom effectively renormalizes the restoring force of the lattice mode. Assuming that the derivative $\partial \Delta_{\mathbf K} / \partial Q_{\mathbf K}$ is positive and varies slowly in time, it may be approximated as a constant set by the electronic susceptibility. The lattice dynamics then correspond to harmonic oscillations with a renormalized frequency
\begin{equation}
	\label{eq:renorm_Omega}
    \tilde{\Omega} = \sqrt{\Omega^2 -\frac{g}{m}\frac{\partial n_{\mathbf{K}}}{\partial Q_{\mathbf{K}}}}.
\end{equation}
This expression makes explicit that the coupling to the electronic subsystem leads to a softening of the lattice mode, $\tilde{\Omega} < \Omega$, consistent with the numerical results shown in Fig.~\ref{K0}(b). 
While this argument provides a useful physical picture, it is necessarily heuristic, as it neglects retardation effects and the full time dependence of the electronic response. A more rigorous analysis of the frequency renormalization, incorporating the dynamical electron susceptibility, will be presented in the following section.

\begin{figure}[t]
\centering
\includegraphics[width=86mm]{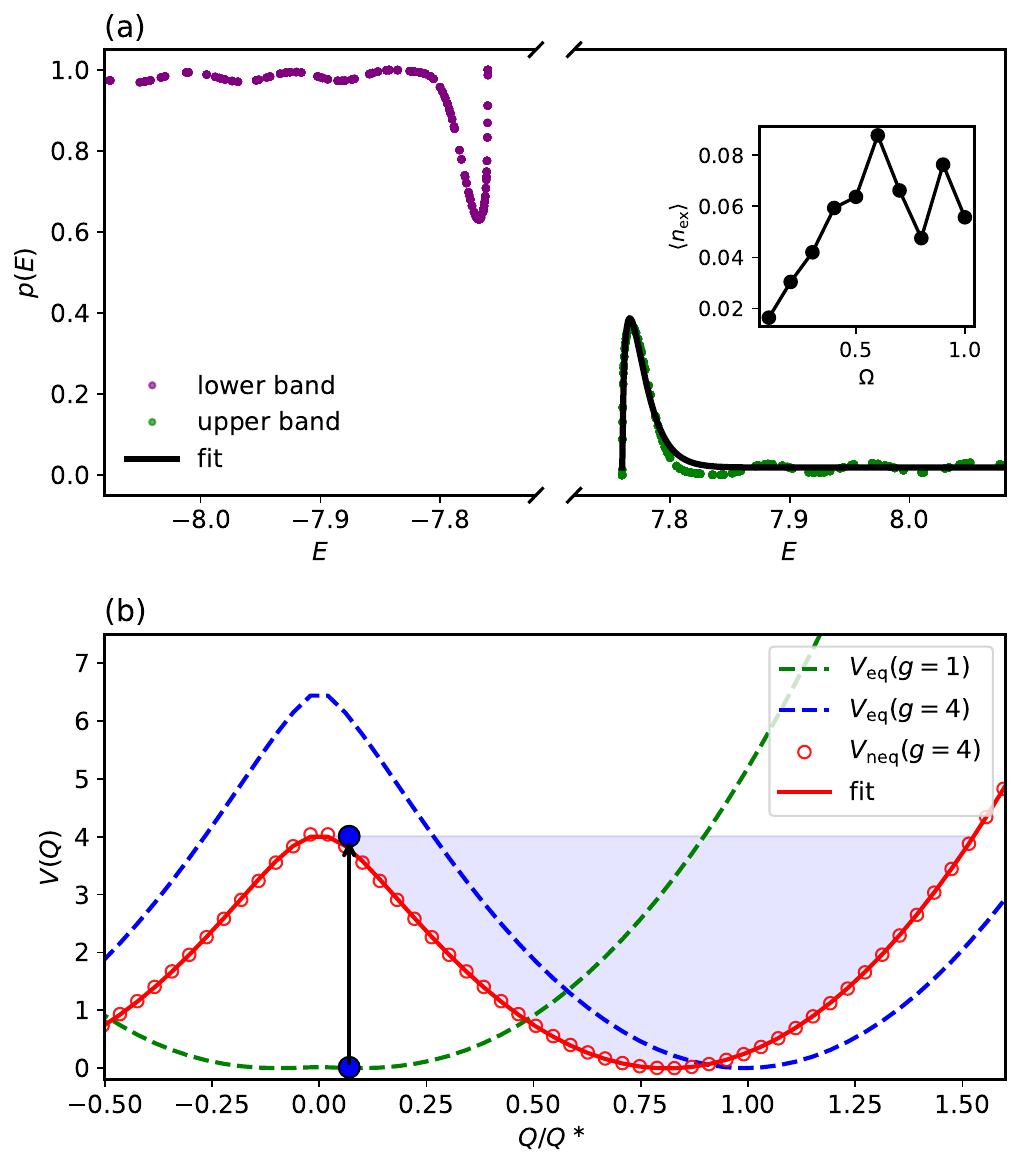}
\caption{
(a) Time-averaged band population over one $Q$ oscillation cycle at
$\Omega = 0.6$. A finite non-equilibrium occupation is visible in
the upper band; the inset shows the corresponding average excited
population \(\langle n_{\mathrm{ex}}\rangle = \int_{\rm RBZ}
\frac{d^{2}p}{(2\pi)^{2}}\; p_{+}(\mathbf{p})\) as a function of phonon
frequency $\Omega$.
(b) Effective potential $V_{\rm eff}(Q_{\mathbf K})$ obtained using the extracted
non-equilibrium population, compared with the equilibrium potentials for
$g=1$ and $g=4$.
}

\label{fig:population}
\end{figure}

\section{Non-equilibrium population and effective potential}
\label{sec:potential}

To develop a physical picture for the dynamics observed here, it is useful to
interpret the collective lattice mode with amplitude $Q_{\mathbf K}$ as evolving
in an effective potential landscape. In equilibrium, this potential is obtained
from the electronic energy evaluated in a static CDW background together
with the bare lattice contribution,
\begin{equation}
\begin{aligned}
V_{\rm eq}(Q)
&=
\langle \hat{\mathcal{H}}_{\rm eL}(Q_{\mathbf K})\rangle
+
\frac{1}{2}K Q_{\mathbf K}^{2}
\\
&=
-\frac{1}{N}
\sum_{\mathbf{p}}
\sqrt{g^{2}Q_{\mathbf K}^{2}+\epsilon_{\mathbf{p}}^{2}}
+
\frac{1}{2}KQ_{\mathbf K}^{2}.
\end{aligned}
\end{equation}
In the non-equilibrium case relevant here, however, the population in the
two instantaneous bands becomes non-thermal due to a quench from $g_i$ to $g_f$.
Therefore, before constructing a corresponding non-equilibrium potential
surface, we first characterize the electronic occupation structure generated
by this quench.

Under the assumption of translation invariance after quench, at each momentum $\mathbf{p}$ the problem remains a two–level system with
instantaneous eigenvalues $\pm \epsilon_{\mathbf{p}}(Q_{\mathbf K})$ and normalized eigenvectors
$|\psi_{\mathbf{p}}^{\pm}(Q_{\mathbf K})\rangle$, where $Q_{\mathbf K}(t)$ evolves slowly.
During the time evolution, the electronic Slater determinant state 
$|\Psi(t)\rangle$ can be projected onto these instantaneous
eigenstates, and the associated occupation probabilities are
\begin{equation}
p_{\pm}(\mathbf{p})
=
|\langle\psi_{\mathbf{p}}^{\pm}(Q_{\mathbf K})|\phi_{\mathbf{p}}( t)\rangle|^{2},
\end{equation}
where $|\phi_{\mathbf p}(t) \rangle$ denotes the time-varying single-particle orbital which constitute the Slater determinant. 

Fig.~\ref{fig:population}(a) shows the time-averaged occupation $p_{\pm}(E) = p_{\pm}(\epsilon_{\mathbf p})$ over one oscillation cycle for a representative case of quench into the phase-locked oscillation. A finite upper-band occupation reflects a genuine non-thermal population generated by the quench dynamics, and is
well approximated by the empirical form
\begin{equation}
p_{+}(E)
=
A(E-E_{\rm min})^{\alpha}
\exp\!\left[-\frac{E-E_{\rm min}}{T}\right]
+
C,
\end{equation}
where $E_{\rm min}$ is the upper–band bottom. The extracted parameters
define a stationary non-equilibrium occupation profile.

Assuming that this occupation remains approximately fixed while the
dispersion $\epsilon_{\mathbf{p}}^{\pm}(Q_{\mathbf K})$ evolves with the slowly varying $Q_{\mathbf K}(t)$, we may construct an effective non-equilibrium
potential for the phonon mode as
\begin{equation}
\begin{aligned}
V_{\rm eff}(Q) &= \langle \hat{\mathcal{H}}_{\rm eL}(Q_{\mathbf K} )\rangle_{\rm neq} + \frac{1}{2}KQ_{\mathbf K}^{2} \\
& =
\sum_{\mathbf{p},\nu=\pm}
p_{\nu}(\mathbf{p})\,\epsilon_{\mathbf{p}}^{\nu}(Q_{\mathbf K})
+
\frac{1}{2}KQ_{\mathbf K}^{2}.
\end{aligned}
\end{equation}
The resulting potential, together with its equilibrium counterparts,
is shown in Fig.~\ref{fig:population}(b). It is well described by the form
\begin{equation}
V_{\rm eff}(Q)
\simeq
\frac{1}{2}K_{\rm eff}Q_{\mathbf K}^{2}
-
\sqrt{g_{\rm eff}^{2}Q_{\mathbf K}^{2}+E_{\rm eff}^{2}}
+
E_{0},
\end{equation}
illustrating that the non-equilibrium population qualitatively reshapes the
energy landscape compared to equilibrium. For the quenches of interest,
the equilibrium potential has a well-defined minimum, whereas the
non-equilibrium distribution lowers and distorts the landscape. Thus, a
quench from $g_i$ to $g_f$ places the system at an energetically displaced
point on this modified surface, naturally generating both a shift in the
preferred value of $Q_{\mathbf K}$ and a finite initial oscillation amplitude $\Delta Q$.

The subsequent dynamics of $Q_{\mathbf K}(t)$ may be interpreted as motion in the
effective anharmonic potential discussed above. A complementary and
more dynamical perspective is obtained directly from the equation of
motion for the lattice mode,
\begin{equation}
m\frac{d^2Q_{\mathbf K}}{dt^2}
=
-\frac{dV_{\rm eff}(Q_{\mathbf K})}{dQ_{\mathbf K}}.
\end{equation}
In the limit $g_f \approx g_i$, the trajectory remains
confined near the minimum $Q_0$ of $V_{\rm eff}(Q)$. Expanding the potential to quadratic
order,
\begin{equation}
V_{\rm eff}(Q_{\mathbf K})\simeq V(Q_0)
+
\frac{1}{2}V_{\rm eff}''(Q_0)(Q_{\mathbf K}-Q_0)^2,
\end{equation}
The oscillation around the minimum is thus characterized by a frequency 
\begin{equation}
\tilde{\Omega} =
\sqrt{\frac{V''(Q_0)}{m}}.
\end{equation}
This curvature-based frequency is fully consistent with Eq.~(\ref{eq:renorm_Omega}) based on the force-based
analysis in the phase-locked regime in the previous section.
The local curvature $V''(Q_0)$ of the non-equilibrium potential therefore
encodes the same physics as the
$g\,\partial n_{\mathbf{K}}/\partial Q_{\mathbf{K}}$ correction. In this weak-quench regime the
motion is essentially harmonic and the oscillation period is amplitude
independent.

For large $|g_i - g_f|$, the dynamics no longer remain restricted to the
harmonic neighborhood of the minimum, and the full anharmonic structure
of $V_{\rm eff}(Q)$ becomes relevant. In this case the restoring force is no longer
linear in $Q$, the curvature-based expression for $\tilde{\Omega}$ ceases
to apply, and the motion must be viewed as nonlinear oscillation between
classical turning points $Q_{\pm}$ where $\dot{Q}=0$. Equivalently, the
oscillation period may be characterized through 
\begin{equation}
\mathcal{T}(E^{\ast})
=
\sqrt{2m}
\int_{Q_-}^{Q_+}
\frac{dQ}{
\sqrt{E^{\ast}-V_{\rm eff}(Q)}
},
\end{equation}
showing explicitly that the oscillation period becomes dependent on the
excitation energy (or amplitude). This amplitude dependence directly
reflects the non-parabolic character of the non-equilibrium potential
surface and represents a genuinely nonlinear lattice response induced
by the quench.

Finally, we note that the above discussion is formulated under the assumption that
the phonon-sector energy $E^{\ast}$ remains constant. In a more complete description,
the lattice mode can exchange energy with the electronic (pseudospin) subsystem,
resulting in dissipation and a gradual relaxation of the oscillation amplitude.
Such dynamical energy transfer processes lie beyond the present static effective
potential framework and would require an explicitly time-dependent treatment.

\section{Conclusion and Outlook}
\label{sec:conclusion}

We have analyzed the nonequilibrium evolution of charge-density-wave order in the half-filled spinless Holstein model following an interaction quench. By focusing on the coherent regime and exploiting the special structure of the CDW order at wave vector $\mathbf K=(\pi, \pi)$, we formulated a pseudospin description that maps the coupled electron–lattice dynamics onto an effective collective-spin problem. Within this framework, the time evolution of the CDW order parameter exhibits a small number of qualitatively distinct dynamical responses that are robust across a broad range of quench parameters. Our simulations identify three characteristic behaviors—locked oscillations, Landau-damped evolution, and overdamped relaxation—which together provide a unified description of the post-quench dynamics in this electron–phonon–coupled system.

While these dynamical regimes bear a strong formal resemblance to those found in quenched BCS superconductors and other electronically driven symmetry-breaking phases, the Holstein model displays an essential qualitative distinction. The presence of dynamical lattice degrees of freedom qualitatively alters the long-time evolution: even in regimes where damping is pronounced, oscillations of the CDW order parameter persist indefinitely rather than decaying completely. We showed that this behavior can be understood in terms of a nonequilibrium electronic distribution that remains coupled to the lattice field, leading to sustained feedback between electronic and phononic sectors. This picture naturally accounts for both the reduced oscillation amplitude and the renormalization of the oscillation frequency relative to the bare Holstein phonon frequency, highlighting the fundamentally hybrid nature of nonequilibrium CDW dynamics in electron–lattice systems.

An important assumption underlying the pseudospin formulation is that the system remains spatially homogeneous after the quench. While this assumption enables a transparent analytical treatment and controlled numerical analysis, spatial homogeneity is not generically guaranteed in interaction quenches, where a substantial amount of energy is suddenly injected into the system. Indeed, the emergence of spatial inhomogeneity---such as domain formation, phase textures, or coexistence of dynamically distinct regions---has been reported in recent studies of quenched superconductors and CDW systems~\cite{chern17,chern19,huang19,yang25,Fan24,Lingyu2024,sankha2024}. Nonetheless, the three dynamical regimes identified within the homogeneous pseudospin framework serve as a useful baseline for understanding more complex nonequilibrium behavior. When spatial fluctuations are allowed, local regions of the system may transiently or persistently realize dynamics analogous to the locked, damped, or overdamped regimes identified here, making the homogeneous theory a natural reference point for interpreting inhomogeneous post-quench states.

Looking forward, our results have direct implications for time-resolved pump–probe experiments on CDW materials with strong electron–phonon coupling. The persistence of coherent oscillations predicted here suggests that long-lived amplitude-mode oscillations may be observable even when strong damping is expected based on purely electronic considerations. Moreover, the renormalization of the oscillation frequency and its dependence on the nonequilibrium electronic distribution provide experimentally accessible signatures in ultrafast reflectivity, time-resolved ARPES, or X-ray scattering measurements. Extending the present framework to include spatially resolved dynamics and disorder may further bridge the gap between theory and experiment, enabling a systematic understanding of how coherent lattice-driven oscillations coexist and compete with spatially inhomogeneous relaxation processes in nonequilibrium CDW systems.

\begin{acknowledgments}
This work was supported by the Owens Family Foundation. Y.Y. was partially supported by the US Department of Energy Basic Energy Sciences under Contract No.~DE-SC0020330. The authors acknowledge Research Computing at The University of Virginia for providing computational resources and technical support.
\end{acknowledgments}

\appendix
\section{Dimensional analysis}
\label{app:dimensional}

For numerical convenience and clarity, we recast the coupled equations of motion for the electronic and lattice degrees of freedom into a dimensionless form. We choose the nearest-neighbor hopping amplitude $t_{\rm nn}$ as the fundamental energy scale and introduce the associated electronic timescale
\begin{equation}
    \tau_e = {\hbar}/{t_{\rm nn}} ,
\end{equation}
which sets the natural unit of time for the electronic dynamics. Dimensionless time is then defined as
$\tilde{t} = t / \tau_e = (t_{\rm nn}/\hbar), t$, and all energies are measured in units of $t_{\rm nn}$. In particular, the single-particle dispersion is rescaled as $\tilde{\epsilon}_{\mathbf p} = \epsilon_{\mathbf p} / t_{\rm nn}$.

As discussed in the main text, a characteristic scale for the lattice distortion can be identified by balancing the elastic energy $K Q^2$ against the electron–lattice coupling energy $g n Q$, yielding
\begin{eqnarray}
	Q^* = g/K,
\end{eqnarray}
We therefore introduce dimensionless lattice variables $\tilde{Q}_i = Q_i / Q^*$. In terms of the rescaled time $\tilde{t}$ and lattice displacement $\tilde{Q}_i$, the von Neumann equation for the electronic density matrix, Eq.~(\ref{eq:von-neumann}), takes the form
\begin{equation}
    i \frac{d\rho_{ij}}{d\tilde{t}}
        = \sum_k \bigl( \rho_{ik} \tilde{t}_{kj} - \tilde{t}_{ik} \rho_{kj} \bigr)
          + 4\lambda \bigl(\tilde{Q}_j - \tilde{Q}_i\bigr)\rho_{ij} ,
    \label{eq:A2}
\end{equation}
where $\tilde{t}_{ij} = t_{ij}/t_{\rm nn}$ and $\lambda$ is the dimensionless electron–phonon coupling,
\begin{equation}
    \lambda = {g^2}/{KW}
\end{equation}
with $W = 4 t_{\rm nn}$ denoting the electronic bandwidth. The parameter $\lambda$ thus controls the strength of electron–lattice feedback in the dimensionless equations of motion.

For the lattice sector, the natural timescale is set by the bare phonon frequency $\Omega$, or equivalently
\begin{eqnarray}
	\tau_L = 1/\Omega = \sqrt{m/K}.
\end{eqnarray}
The ratio of lattice and electronic timescales defines the adiabatic parameter
\begin{equation}
    r = \tau_L / \tau_e =  {\hbar \Omega}/{t_{\rm nn}},
\end{equation}
which controls the relative speed of lattice motion compared to electronic dynamics. A characteristic momentum scale for the lattice follows from the relation $dQ/dt = P/m \sim \Omega Q$, giving
\begin{eqnarray}
	P^* = m \Omega Q^*
\end{eqnarray}
Introducing the dimensionless momentum $\tilde{P}_i = P_i / P^*$, the Newton equations of motion for the lattice degrees of freedom can be written in dimensionless form as
\begin{eqnarray}
	\frac{d\tilde{Q}_i}{d \tilde{t}} = r \tilde{P}_i, \qquad 
	\frac{d\tilde{P}}{d \tilde{t}} = -r \tilde{Q}_i + r\left( \rho_{ii} -\frac{1}{2} \right).
\end{eqnarray}
Finally, expressing the electronic density matrix in terms of the pseudospin components $S_{\mathbf p}^\alpha$ as introduced in Sec.~\ref{sec:model}, and using the dimensionless time $\tilde{t}$ and staggered lattice distortion $\tilde{Q}_{\mathbf K}$, we obtain the pseudospin equations of motion in fully dimensionless form:
\begin{subequations}
\begin{align}
    \frac{d S_{\mathbf p}^z}{d\tilde{t}} &= 8\lambda\,\tilde{Q}_{\mathbf K}\, S_{\mathbf p}^y, \\
    \frac{d S_{\mathbf p}^x}{d\tilde{t}} &= 2\tilde{\epsilon}_{\mathbf p}\, S_{\mathbf p}^y,   \\
    \frac{d S_{\mathbf p}^y}{d\tilde{t}} &= -2\tilde{\epsilon}_p\, S_{\mathbf p}^x
                                - 8\lambda\,\tilde{Q}_{\mathbf K}\, S_{\mathbf p}^z.  
\end{align}
\end{subequations}
These equations make explicit that the post-quench dynamics is governed by two dimensionless control parameters, $\lambda$ and $r$, which encode the strength of electron–phonon coupling and the relative separation of electronic and lattice timescales, respectively.

\section{Asymptotic behavior for quench $g_i > 0 \rightarrow g_f = 0$}

\label{app:zero-gf}

\begin{figure}
    \centering
    \includegraphics[width=0.85\linewidth]{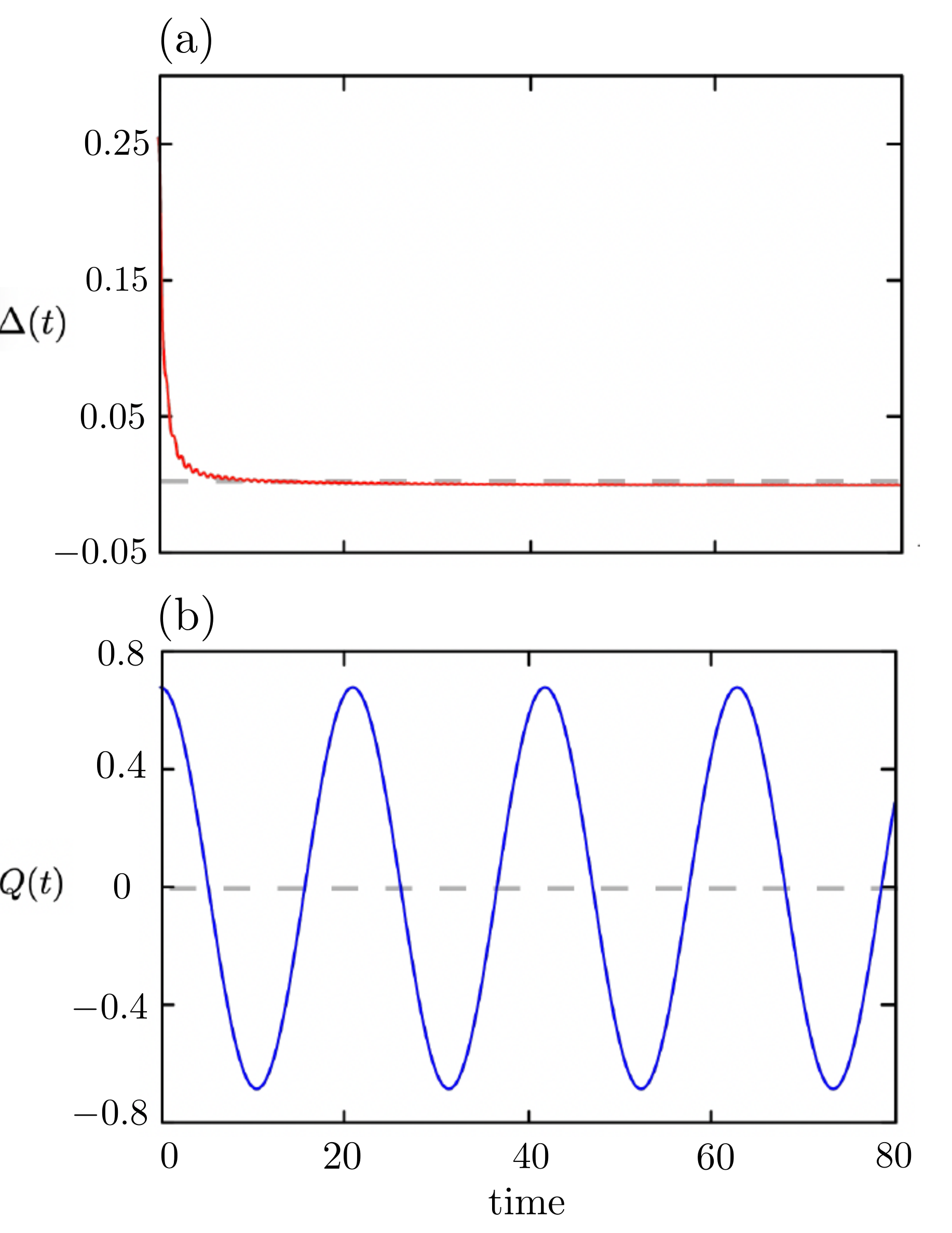}
    \caption{Post-quench dynamics in the decoupled limit $g_f=0$. (a) The CDW order parameter $\Delta(t)$ rapidly dephases and decays toward zero as contributions from different momenta interfere destructively. (b) The lattice coordinate $Q(t)$ is decoupled from the electronic sector and exhibits undamped harmonic oscillations.}
    \label{perturbation_g=0}
\end{figure}

In the limit $g_f=0$, the post-quench electron–phonon interaction vanishes, and the lattice degrees of freedom decouple entirely from the electronic sector, undergoing simple harmonic motion of the single relevant phonon mode. The electronic dynamics are then governed solely by the Bloch equation [Eq.~\eqref{eq:Bloch_eq}], with each pseudospin precessing about a fixed, time-independent effective field. The resulting dynamics admit an exact solution, which can be written as
\begin{equation}
\begin{aligned}
S_{\mathbf p}^{x}(t) &= \sin(\alpha_{\mathbf p}) \cos(2i\epsilon_{\mathbf p} t), \\
S_{\mathbf p}^{y}(t) &= \sin(\alpha_{\mathbf p}) \sin(2i\epsilon_{\mathbf p} t), \\
S_{\mathbf p}^{z}(t)    &= \cos(\alpha_{\mathbf p}), \\
Q_{\mathbf K}(t)   &= Q^{(i)}_{\mathbf K}\cos(\Omega t),
\end{aligned}
\end{equation}
where the pseudospin length has been normalized to unity and $\Omega$ denotes the bare phonon frequency. The angle $\alpha_{\mathbf p}$ is the initial polar angle (in the $xz$ plane) between the pseudospin at momentum $\mathbf p$ in the pre-quench ground state and the $z$ axis. It is determined by the initial coupling $g_i$ according to
\begin{equation}
\tan \alpha_{\mathbf p}=-\frac{g_i Q^{(i)}_{\mathbf K}}{\epsilon_{\mathbf p}},
\end{equation}
Here $Q_{\mathbf K}^i$ denotes the initial equilibrium amplitude of the $\mathbf K = (\pi,\pi)$-mode lattice distortion. 

In the thermodynamic limit, the CDW order parameter $\Delta_{\mathbf K}(t)$ is obtained by integrating the pseudospin coherence over momentum space within the reduced Brillouin zone (RBZ),
\begin{equation}
\Delta_{\mathbf K}(t)
= \mathrm{Re}\!\left\{\int_{\mathrm{RBZ}} \frac{d^{2}p}{(2\pi)^{2}}\,
\sin(\alpha_{\mathbf p})\, e^{2 i \epsilon_{\mathbf p} t}\right\}.
\end{equation}
The long-time behavior of $\Delta_{\mathbf K}(t)$ follows from a stationary-phase analysis of the momentum integral. The dominant contribution arises from stationary points of the phase, defined by $\nabla_{\mathbf p}\epsilon_{\mathbf p}=0$. Within the RBZ, this condition is satisfied at $\mathbf p=\mathbf 0$. Expanding the dispersion to quadratic order about this point and retaining the leading contribution to the phase, the order parameter reduces asymptotically to
\begin{equation}
\Delta_{\mathbf K}(t)\sim \int_{|\mathbf p|<\Lambda} d^{2}p\, e^{-i t p^{2}},
\end{equation}
where $\Lambda$ is a momentum cutoff. For finite $\Lambda$, this oscillatory Gaussian integral can be evaluated in closed form and yields an envelope that decays as $1/t$. Consequently, the leading asymptotic behavior is 
\begin{eqnarray}
	\Delta_{\mathbf K}(t)\sim t^{-1}, 
\end{eqnarray}
up to an overall prefactor and phase. This algebraic decay reflects dephasing: the momentum dependence of $\epsilon_{\mathbf p}$ produces a continuum of pseudospin precession frequencies, leading to the gradual suppression of the macroscopic order parameter. This behavior is illustrated in Fig.~\ref{perturbation_g=0}(a).

\bibliography{ref}
\end{document}